\documentclass[a4paper,twocolumn,showpacs,amsmath, amssymb,nofootinbib,aps,floatfix]{revtex4}
\usepackage{epsfig}
\input psfig.sty 

\usepackage{bm}									
\usepackage{dcolumn}								
\usepackage{graphicx}								
\usepackage{hyperref}

\def \be {\begin{equation}} 

\def \ee {\end{equation}} 
\def \bea {\begin{eqnarray}} 
\def \eea {\end{eqnarray}} 

\usepackage{graphicx}
\usepackage{dcolumn}
\usepackage{bm}
\usepackage{epsfig}

\newcommand*{\ltsim}{\ {\raise-.75ex\hbox{$\buildrel<\over\sim$}}\ }
\newcommand*{\gtsim}{\ {\raise-.75ex\hbox{$\buildrel>\over\sim$}}\ }
\newcommand*{\proptosim}{\ {\raise-.75ex\hbox{$\buildrel\propto\over\sim$}}\ }

\begin{document}
\title{Cosmological constraints on the gas depletion factor in galaxy clusters} 
\author{R. F. L. Holanda$^{1,2}$} \email{holanda@uepb.edu.br}
\author{V. C. Busti$^{3,4}$}\email{viniciusbusti@gmail.com}
\author{J. E. Gonzalez$^{5}$}\email{javierernesto@on.br}
\author{F. Andrade-Santos$^{5}$} \email{fsantos@cfa.harvard.edu}
\author{J. S. Alcaniz$^{6,7,8}$}\email{alcaniz@on.br}
\affiliation{$^1$Departamento de F\'{\i}sica, Universidade Federal de Sergipe, 49100-000, Sao Cristovao - SE, Brasil}
\affiliation{$^2$Departamento de F\'{\i}sica, Universidade Federal de Campina Grande, 58429-900, Campina Grande - PB, Brasil}
\affiliation{$^3$Department of Physics and Astronomy, University of Pennsylvania, Philadelphia, PA 19104, USA} 
\affiliation{$^4$Departamento de F\'{\i}sica Matem\'{a}tica, Universidade de S\~{a}o Paulo, 
CEP 05508-090, S\~{a}o Paulo - SP, Brasil}
\affiliation{$^5$Harvard-Smithsonian Center for Astrophysics, 60 Garden Street, Cambridge, MA 02138, USA}
\affiliation{$^6$Observat\'orio Nacional, 20921-400, Rio de Janeiro - RJ, Brasil}
\affiliation{$^7$Departamento de F\'\i sica, Universidade Federal do Rio Grande do Norte, 59072-970 Natal, RN, Brasil}
\affiliation{$^8$Physics Department, McGill University, Montreal, QC, H3A 2T8, Canada}

\date{\today}

\begin{abstract}

The evolution of the X-ray emitting gas mass fraction ($f_{gas}$) in massive galaxy clusters  can be used as an independent  cosmological tool to probe the expansion history of the Universe. Its use, however, depends upon a crucial quantity, i.e., the  depletion factor $\gamma$, which corresponds to  the ratio by which $f_{gas}$ is depleted with respect to the universal baryonic mean. This quantity is not directly observed and hydrodynamical simulations performed in a  specific cosmological model (e.g., a flat $\Lambda$CDM cosmology) have been used to calibrate it. In this work, we obtain for the first time self-consistent observational constraints on the gas depletion factor combining 40 X-ray emitting gas mass fraction measurements  and luminosity distance measurements from  type Ia supernovae. Using Gaussian Processes to reconstruct a possible redshift evolution of $\gamma$, we find no evidence for such evolution, which confirms the current results from hydrodynamical simulations. {  Moreover, our constraints on $\gamma$ can be seen as a data prior for cosmological analyses on different cosmological models. The current measurements are systematic limited, so future improvements will depend heavily on a better mass calibration of galaxy clusters and their measured density profiles.}
  
\end{abstract}
\maketitle

\section{Introduction}

Currently, cosmological observations are able to constrain the main cosmological parameters within a few percent, as well as to test the observational viability of a number of cosmological models. These results are mainly obtained from a combination of high precision measurements of the temperature fluctuations of the cosmic microwave background (CMB) \cite{ade}, the imprint of the baryon acoustic oscillations (BAO) in the clustering of galaxies~ \cite{ein,wei,gab}, and observations of hundreds of type Ia supernovae (SNe Ia) at low and intermediary redshifts \cite{suz,jla}. Together, these observables also led to the establishment of the $\Lambda$CDM model as the standard cosmology, whose the values of its main parameters were recently summarized by the {{\it Planck} Collaboration} \cite{ade}. 

The aforementioned data have also been used, together with other observables, to test fundamental hypotheses of the standard cosmological model, as the validity of the assumption of homogeneity and isotropy of the Universe on large scales (see e.g. \cite{cla} and references therein), the constancy of the fine structure constant \cite{webb1,webb2,hol1,hol2}, and the validity of the cosmic distance duality relation (CDDR), which relates the luminosity distance $D_L$ of an object to its angular diameter distance $D_A$ as $D_L/D_A(1+z)^{2}=1$ \cite{eth,eli}. Currently, the tightest  constraints on the CDDR to date come from the blackness of the CMB spectrum, which requires that the above relation cannot be violated by more than $0.01\%$ from decoupling until today \cite{esc}, and from measurements of the gas mass fraction of massive galaxy clusters from Sunyaev-Zel'dovich and X-ray observations \cite{hol3,hol4}.

In particular, these massive clusters are interesting tools for cosmology since their baryon content is expected to trace closely the cosmic baryon content, $\Omega_b$ (the ratio of the baryon density $\rho_b$ to the critical density) \cite{sas}. By assuming that the measurements of the X-ray emitting gas mass fraction do not evolve with redshift, this quantity has been  used  to constrain  the geometry of the universe, the  matter (baryonic plus dark) density parameter $\Omega_M$ and the dark energy equation of state $w$ \cite{lima2003,allen1,allen2,allen3,allen4,ett1,ett2,mantz}. However, it is worth mentioning that present cosmological constraints from X-ray emitting gas mass fraction  observations depend  on hydrodynamical simulations \cite{bat,pla}. This in turn has been used to link the observed X-ray emitting gas mass fraction (henceforward  gas mass fraction) to the cosmic baryon fraction, with the extra factor being the so-called  depletion factor, i.e., $\gamma=f_{gas}(\Omega_b/\Omega_M)^{-1}$, which in principle may be a function of redshift.\footnote{Current optical and X-ray observations at low redshifts seem to indicate a baryon fraction in clusters which is smaller than expected \cite{ett2}, giving rise to different explanations, such as undetected baryon components \cite{ett1,lin}, underestimation of $\Omega_{\rm M}$ by CMB probes \cite{ade}, among others \cite{bia}.} 

Measuring the amount  of the gas mass fraction ($f_{gas}$)  and its possible evolution with redshift it is crucial for a better  understanding of the galaxy cluster physics. Nowadays, there are improved hydrodynamical simulations of galaxy cluster formation that take into account a realistic amount of energy feedback from active galaxy nucleus and supernovae in addition to radiative cooling and star formation.  {In this kind of approach the depletion factor is usually parametrized by an arbitrary  function of the redshift $z$, such as $\gamma(z)=\gamma_0(1+\gamma_1 z)$. The Refs.\cite{pla,bat} considered the  gas mass fraction as a cumulative quantity into $r_{2500}$, the  radii at which the mean cluster density is 2500 times the critical density of the Universe at the cluster's redshift, and obtained the intervals $0.55 \leq \gamma_0 \leq 0.79$ and $-0.04 \leq \gamma_1 \leq 0.07$, depending on the physical processes that are included in simulations \footnote{Recent simulations exploring the global properties and hot gas profiles  into $r_{200}$ of clusters  at low-redshift can be found in Ref.\cite{barnes}.} (see Table 3 of \cite{pla}). Therefore, no significant evolution with redshift has been verified. However, it is important to mention that these hydrodynamic simulations considered a flat $\Lambda$CDM model as the background scenario, with $\Omega_M=0.24$, $\Omega_b = 0.04$, $H_0 = 72$ km/s/Mpc and the primordial spectral index and normalization of the power spectrum given, respectively, by $n_s = 0.96$ and  $\sigma_8 = 0.8$. Moreover, by comparing their results from radiative simulations for stellar fraction in massive galaxy clusters with observations, the authors of Ref.\cite{pla} found a larger stellar fraction in massive galaxy clusters, independent of the observational data used in comparison (see Fig. 2 in their paper). In principle, this may occur due to difficulty in distinguishing in simulations the stars in the diffuse stellar component and in the central galaxy, however, it also possible that the physical processes used in hydrodynamic simulations do not span the entire range of physical processes allowed by our current understanding of the intra-cluster medium.}

{On the other hand, the results from simulations for spherical shells at radii near $r_{2500}$ ($0.8 < r/r_{2500} < 1.2$) showed that the $\gamma_0$ value  presents  only a slightly dependence on physical processes. In such spherical shells the stellar contribution can be negligible and $\gamma$ is constrained to be $\gamma_0= 0.85 \pm 0.08$ (see Fig.6 in \cite{mantz} and \cite{pla}). However, available information from the  hydrodynamical simulations is insufficient  to obtain a well-motivated prior on $\gamma_1$ for gas mass fraction measurements in such shells. This occurs because the authors of Ref.\cite{pla} obtained the $\gamma$ values for spherical shells at radii near $r_{2500}$, $0.8 < r/r_{2500} < 1.2$, only at $z=0$ and $z=1$. So, a conservative prior was  adopted in \cite{mantz} ($-0.05 \leq \gamma_1 \leq 0.05$) to derive constraints on cosmological parameters.}

In this paper, we take a different approach. Assuming the validity of the CDDR, we use cosmological observations, such as 40 gas mass fraction measurements in galaxy clusters \cite{mantz} and luminosity distances of  type Ia supernovae  \cite{suz,jla}, to explore the behavior of the gas depletion factor up to redshift one. Unlike previous works, no specific cosmological model is considered in the analyses \footnote{Recently, a similar approach was performed in Ref.\cite{hol44} to put constraints on a possible evolution of mass density power-law index in strong gravitational lensing. By considering the CDDR validity, SNe Ia and strong gravitational lensing systems they obtained a mild evolution for the power-law index.}. By using Gaussian Processes (GPs) to reconstruct a possible redshift evolution of $\gamma$, we find a very good agreement with the aforementioned results from hydrodynamic simulations. {  Finally, we also consider the 40 gas mass fraction measurements in galaxy clusters and luminosity distances for each one of them obtained from the flat $\Lambda$CDM model constrained by the current CMB experiments \cite{ade}. By adopting a simple function for $\gamma(z)$, $\gamma(z)=\gamma_0(1+\gamma_1z)$, the results are in full agreement with those from SNe Ia data with no evidence for redshift evolution of the $\gamma(z)$.}

The paper is organized as follows. In Sec. II we introduce the basic theoretical background used in the analyses. Sec. III describes the samples used in the statistical analyses. Sec. IV presents the results and Sec. V
a discussion. We lay out our conclusions in Sec. VI.

\section{Theoretical background}

In this section, we present the  theoretical background  used in our method to reconstruct a possible redshift evolution of $\gamma$ parameter. We discuss the cosmic distance duality relation, the gas mass fraction and the Gaussian processes.

\subsection{The cosmic distance duality relation}

 The CDDR  relates the luminosity distance $D_L$ of an object to its angular diameter distance $D_A$ as $D_L/D_A(1+z)^{2}=1$. Actually, it is the astronomical version of the reciprocity theorem proved long ago in Ref.\cite{eth} and requires only that source and observer are connected by null geodesics in a Riemannian spacetime and the  number of photons  conservation (see also \cite{eli}). Although a number of analysis have recently tried to establish whether or not the CDDR holds in practice using observational data, the majority of the studies in observational and theoretical cosmology assume this expression to be valid. We will adopt the latter approach since the expected deviations from this relation are very small when compared to the current observational uncertainties (see, e.g. Table I of \cite{hol5} for a summary of recent analyses involving several astronomical observations).

\subsection{The gas mass fraction}

 {  The cosmic gas mass fraction is defined as $f_{gas}=\Omega_b/\Omega_M$. The assumed constancy of this quantity within massive, relaxed clusters can be used to constrain cosmological parameters following expression \cite{allen3,mantz}:}
\begin{equation} \label{eq:fgasmodel}
  f_{gas}^\mathrm{ref}\left(z\right) = K(z) \, A(z)\, \gamma(z) \left( \frac{\Omega_b}{\Omega_M} \right) \left[ \frac{D_A^\mathrm{ref}(z)}{D_A(z)} \right]^{3/2},
\end{equation}
where 
\begin{equation} \label{eq:thetarat}
  A(z) = \left( \frac{\theta^\mathrm{ref}_{2500}}{\theta_{2500}} \right)^\eta \approx \left( \frac{H(z) \, D_A(z)}{\left[H(z) \, D_A(z)\right]^\mathrm{ref}} \right)^\eta.
\end{equation}
Using the CDDR, one may solve for $\gamma(z)$ to obtain
\begin{equation} \label{gamma}
\gamma(z)=\left(\frac{H(z)^\mathrm{ref}}{H(z)}\right)^{\eta}\frac{f_{gas}^\mathrm{ref}}{K(\Omega_b/\Omega_M)}\left(\frac{D_L}{D_L^\mathrm{ref}}\right)^{3/2 - \eta}.
\end{equation}
The parameters in the above equation are the  following: $K(z)$ quantifies inaccuracies in instrument calibration, as well as any bias in the masses measured due to substructure, bulk motions and/or non-thermal pressure in the cluster gas; the power-law slope $\eta$ has its value averaged over the cluster sample whereas the factor $A(z)$ accounts for the change in angle subtended by $r_{2500}$ as the underlying cosmology is varied (see section 4.2 of \cite{allen3} for details); finally, the index ``ref" corresponds to the fiducial cosmological model used to obtain the $f_{gas}^\mathrm{ref}$ (a flat $\Lambda$CDM model with Hubble constant $H_0=70$ km s$^{-1}$ Mpc$^{-1}$ and the present-day matter density parameter $\Omega_M=0.3$). {  It is important to comment that the ratio into brackets in Eq.(\ref{eq:fgasmodel}) computes the expected variation in $ f_{gas}^\mathrm{ref}\left(z\right)$ when the underlying cosmology is varied. In our case, we consider that the actual cosmology, more precisely, the angular diameter distance $D_A$ in Eq.(1) (or $D_L$ in Eq.(3)) for each galaxy cluster, is given by the SNe Ia data. The term $D_L^\mathrm{ref}$ in Eq.(3) rule out all dependence of the $f_{gas}^\mathrm{ref}$ with respect to the reference cosmological model used in the observations. }

\begin{figure*}
\label{fig_gamma}
\centering
\includegraphics[width=0.47\textwidth]{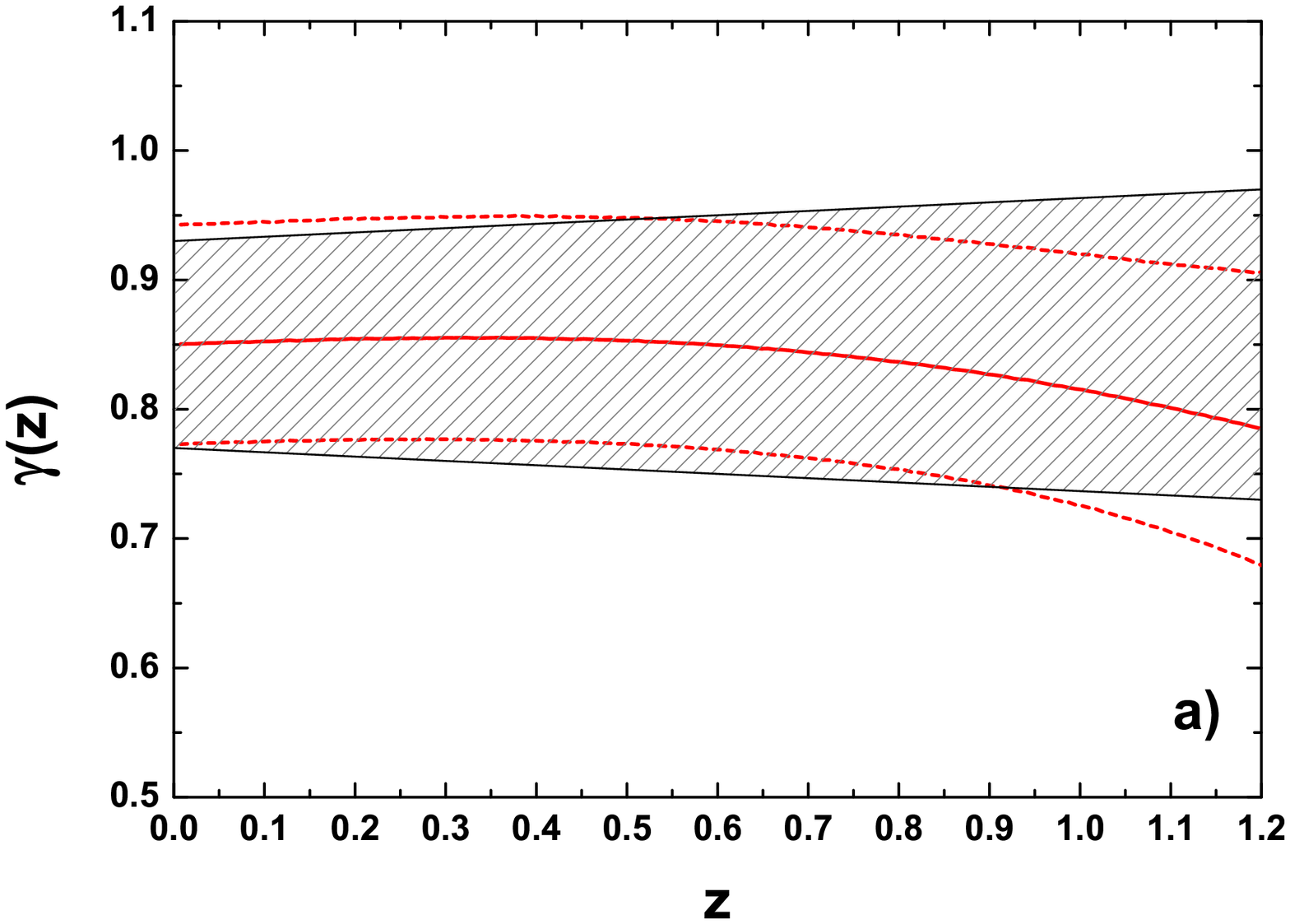}
\includegraphics[width=0.47\textwidth]{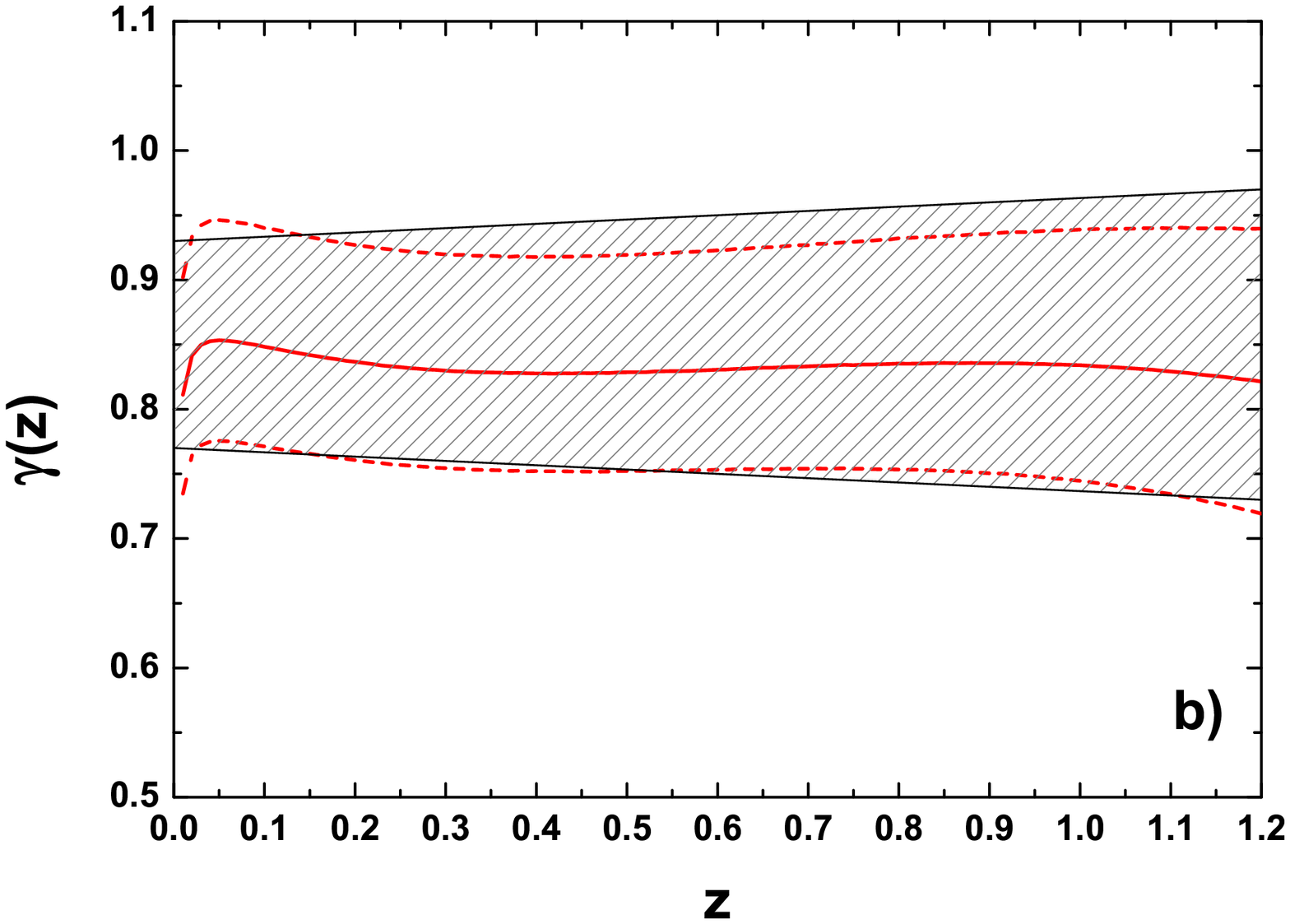}
\caption{Fig.(a) and (b) show the results by using the gas mass fraction measurements {\it plus}  Union2.1 and JLA SNe Ia compilations, respectively. In both figures the blue filled regions correspond to our reconstruction of the gas depletion factor as a redshift  function by using GPs. The hatched  regions in both figures correspond to the results obtained by adopting $\gamma(z)=\gamma_0(1+\gamma_1z)$ with the value for $\gamma_0$ and $\gamma_1$ from the most recent hydrodynamical simulations \cite{pla}.}
\end{figure*}

\begin{figure*}
\label{fig_gamma}
\centering
\includegraphics[width=0.47\textwidth]{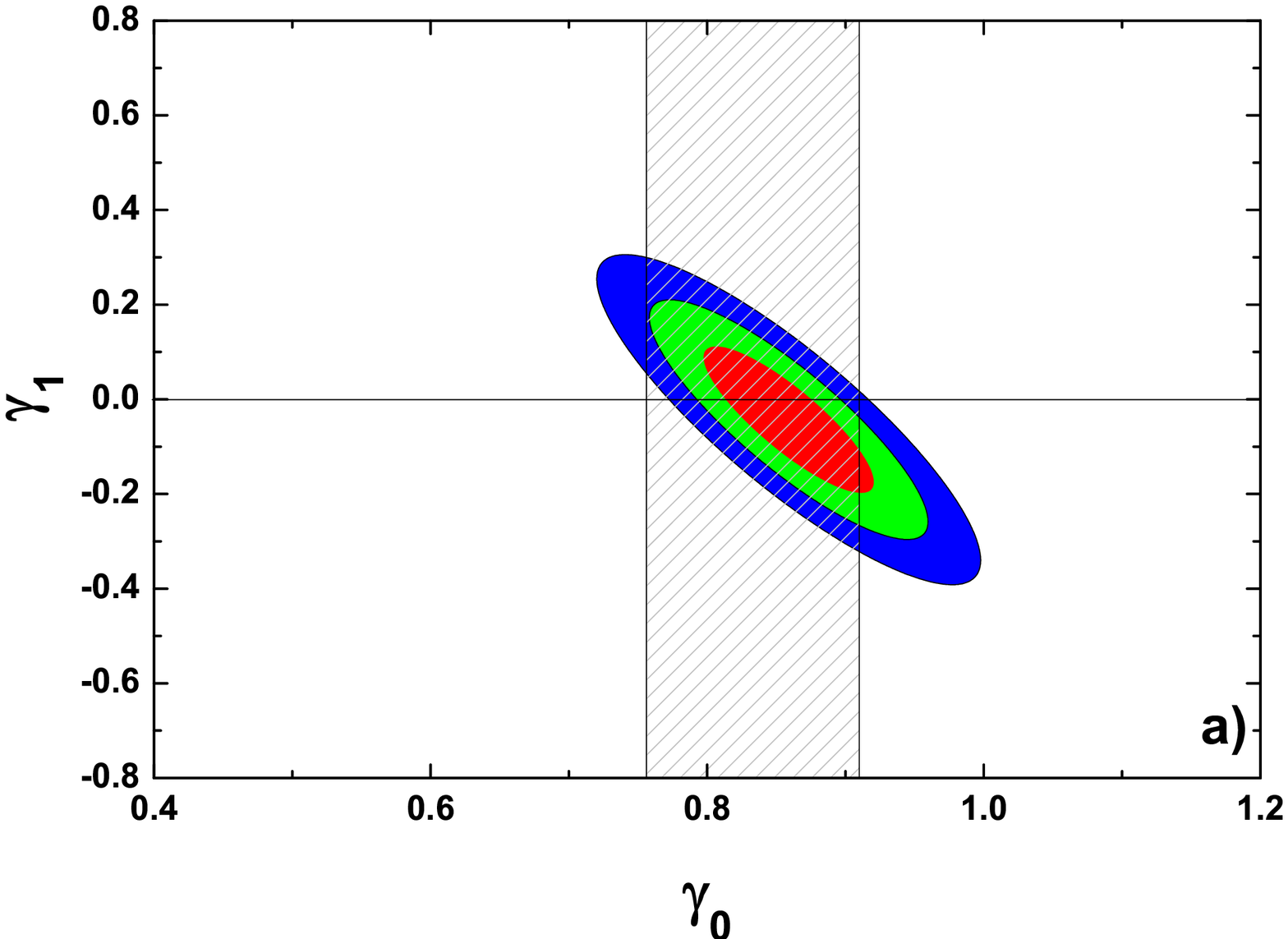}
\includegraphics[width=0.47\textwidth]{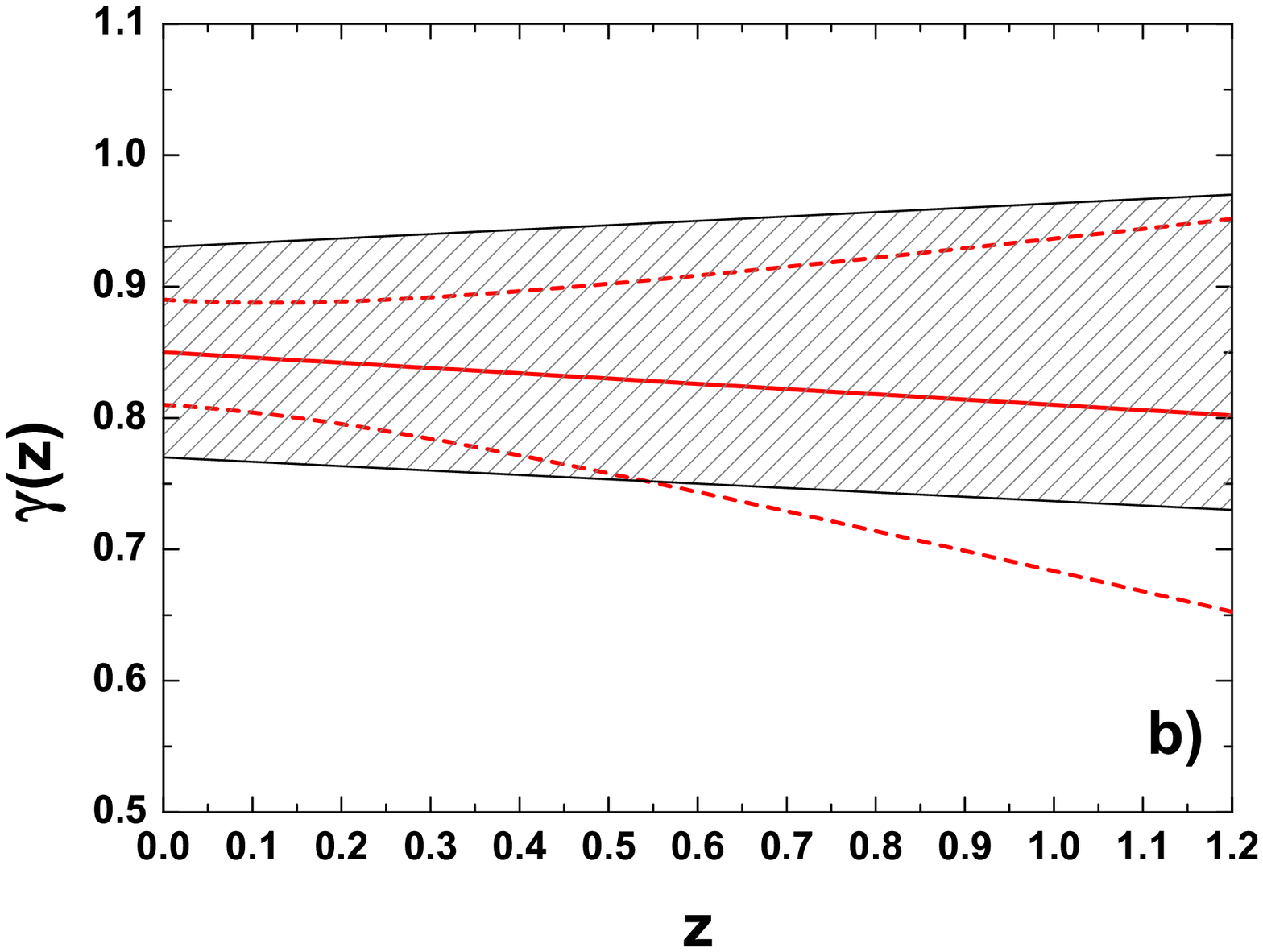}
\caption{Fig.(2a)  shows the 1$\sigma$, 2$\sigma$ and 3$\sigma$ c.l. regions for $\gamma_0$ and $\gamma_1$ by using the luminosity distances given by the current CMB experiments \cite{ade} on the flat $\Lambda$CDM framework. For this case, we consider a simple function for $\gamma(z)$ ($\gamma(z)=\gamma_0+\gamma_1z$). The hatched  region  corresponds to the result obtained by  the most recent hydrodynamical simulations \cite{pla}. Fig.(2b) shows the evolution of the function $\gamma(z)$ and its 1$\sigma$ interval.  }
\end{figure*}

\subsection{Gaussian Processes}

GPs are a generalization of a Gaussian random variable into a Gaussian random function, being characterized by a mean and a covariance function. The covariance function dictates how the function changes in the $x$ or $y$ axes, and also how smooth the process is, that is, how many derivatives can be taken. These features are controlled by hyperparameters, where a common class of covariance functions is the Mat\'{e}rn family:
\begin{equation}
k(z,\tilde{z}) = \sigma_f^2 \frac{2^{1-\nu}}{\Gamma(\nu)} \left [ \frac{\sqrt{2\nu(z-\tilde{z})^2}}{l}  \right]^{\nu} K_{\nu}\left(\frac{\sqrt{2\nu(z-\tilde{z})^2}}{l}  \right).
\end{equation}
In the above equation the hyperparameters are $\sigma_f$, $l$ and $\nu$, where $\sigma_f$ controls changes in the y axis, $l$ in the x axis and $\nu$ the smoothness of the process. $K_{\nu}$ is a modified Bessel function. When $\nu \rightarrow \infty$, one obtains the squared exponential covariance function $k(z,\tilde{z}) = \sigma_f^2 \exp(-(z-\tilde{z})^2/2l^2)$, in which all its derivatives exist and are continuous.  {When we lower the value of $\nu$, less and less derivatives can be taken, up to $\nu = 1/2$, where no derivative can be taken and it is generally used to model Brownian motion.} The hyperparameters should be optimized or marginalized following standard procedures. 

Here, we use GaPP (Gaussian Processes in Python) \cite{gapp} to reconstruct the evolution of $D_L(z)$ or, equivalently,  the normalized comoving distance $D = (H_0/c) D_L/(1+z)$ and its derivative $H(z) = H_0/D^{\prime}(z)$ from SNe Ia distance measurements.   {We adopted $\nu = 9/2$, which had the best coverage properties in extensive simulations performed by \cite{mar}. The hyperparameters $\sigma_f$ and $l$ were optimized through a maximum likelihood method following the steps in Ref.\cite{gapp}. While marginalization of the hyperparameters could in principle provide more robust results, the Ref.\cite{mar} have shown that essentially indistinguishable results are derived for the sample we are considering.} We use the same approach to reconstruct $f_{gas}$ and then obtain $\gamma(z)$ from Eq.(\ref{gamma}),  where we selected the squared exponential covariance function, but we checked no noticeable change was achieved by selecting covariance functions of the Matern family. {  For the JLA, Union2.1 and gas mass fraction samples, the values obtained for $\sigma_f$ and $l$ are, respectively: 12260, 2.25; 100.00, 28.69 and 0.12, 14.62.}

\section{Gas mass fraction and SNe Ia Data}

In order to reconstruct a possible time evolution of the gas depletion factor according to the previous sections, we use the following current data of type Ia supernovae (SNe Ia) and $f_{gas}$ measurements:

\begin{itemize}
\item  580 SNe Ia data compiled by Ref.\cite{suz}, the so-called Union2.1 compilation, with redshift range $0.015 \leq z \leq 1.414$.  The Union2.1 SNe Ia compilation is an update of the Union2 compilation, as stressed by the authors and all SNe  Ia were fitted using SALT2-1 \cite{guy}.  We take into account all the systematic errors in our analysis,  {which are: color correction, mass correction, intergalactic extinction, galactic extinction normalization, rest-Frame U-Band calibration, lightcurve shape, Malmquist Bias, NICMOS Zeropoints, ACS Filter Shift, ACS Zeropoints, all instrument calibration and Vega star magnitude.  Estimates of the systematic error are entered into a covariance matrix. The effect on  constant $\omega$ error, for instance,  where $\omega$ is the dark energy state equation parameter, for each type of systematic error can be found in Table 5 of \cite{suz}.} 
\item We also consider the 31 binned distance modulus  from the JLA compilation and the respective Covariance matrix (see Tables F.1 and F.2 of \cite{jla}). The binned data set is in the redshift range $0.03 \leq z \leq 1.30$. The original data set  includes  740 spectroscopically confirmed SNe Ia with high quality light curves.
\item The galaxy cluster sample is the one reported in Ref.\cite{mantz}. Under the assumptions of spherical symmetry and hydrostatic equilibrium, the data set consists of  40 $f_{gas}$ measurements  in the redshift range $0.078 \leq z \leq 1.063$ observed by the Chandra telescope, identified as massive,  morphologically relaxed systems and with $kT \geq 5 \rm{keV}$. Actually, this sample contains the most dynamically relaxed, massive clusters known. The $f_{gas}$ measurements were taken on a  $(0.8-1.2)$ $\times r_{2500}$ shell rather than  integrated at all radii $r \leq r_{2500}$. This radii is the typical radius within which precise measurements for the $f_{gas}$ have been carried out so far for distant clusters using the Chandra telescope. The exclusion of cluster centers from this measurement significantly reduces the corresponding theoretical uncertainty in gas depletion from hydrodynamic simulations. If compared with previous works, the systematic uncertainties were reduced by incorporating a robust gravitational lensing calibration of the X-ray mass estimates and by restricting the measurements to the most self-similar and accurately measured regions of clusters. The $K(z)$ parameter for this samples was estimated to be  $K=0.96 \pm 0.09$ and no significant trends  with mass, redshift or the morphological indicators were verified \cite{app}. The power-law slope $\eta$ ($0.442 \pm 0.035$) has its value averaged over the cluster sample. We also use  priors on the $\Omega_b$ and $\Omega_M$ parameters, i.e.,  $\Omega_b = 0.0480 \pm 0.0002$ and  $\Omega_M = 0.3156 \pm 0.0091$, as given by current CMB experiments \cite{ade}. {  These priors are from analyses by using exclusively CMB observations on the flat $\Lambda$CDM model.}
\end{itemize}

\section{Results}

Fig.1a shows the result of the reconstruction process by using the $f_{gas}$ measurements and the SNe Ia from Union2.1 compilation \cite{suz}. We also perform our analyses considering the 31 binned distance moduli of SNe Ia from the JLA compilation \cite{jla} and the corresponding covariance matrix. The result is plotted in Fig.1b. In both figures, the blue filled regions correspond to our reconstruction of the gas depletion factor as a redshift  function by using GPs. The hatched  regions  correspond to the results obtained by adopting $\gamma(z)=\gamma_0(1+\gamma_1z)$ with the value for $\gamma_0$ from the most recent hydrodynamical simulations \cite{bat,pla}, $\gamma_0=0.85\pm 0.08$,  and the conservative prior on $\gamma_1$ considered by Ref.\cite{mantz}, $\gamma_1=0.00 \pm 0.05$. In Table I we also show the results of our analysis from both SNe Ia data at different redshifts, i.e., $z = 0$, $z = 0.5$ and $z = 1.0$.

Clearly, the results from both SNe Ia compilations are in full agreement each other and support  no significant evolution of the depletion factor $\gamma$, which is the fundamental hypothesis in the gas mass fraction test. We emphasize that the method proposed here to constrain a possible evolution of the depletion factor $\gamma$ are in line with the arguments implicit in the original papers about the gas mass fraction as a cosmological test \cite{sas,pen}, in that local properties of galaxy clusters can be constrained by a global arguments, in our case provided by the cosmic distance duality relation and SNe Ia observations.

{  Moreover, we can see that the difference by changing from the Union2.1 sample to the JLA one was very small for $\gamma(z)$. Actually, this is due to the fact that the errors are dominated by the priors on the mass calibration and the power-law index, 9.4\% for $K$ and 7.8\% for $\eta$. So, even with higher errors from the reconstructions at higher redshifts since there are fewer objects in that region, errors on $\gamma(z)$ had increased only slightly, from 10.9\% at $z=0$ to 12.5\% at $z=1.0$. Therefore, our results are systematic limited, improvements will come mostly from the understanding of systematic errors.}

{  We also perform the analysis by using the 40 gas mass fraction measurements and  luminosity distances  from the flat $\Lambda$CDM model constrained exclusively by the current CMB observations \cite{ade}. In such model,  the luminosity distance for each galaxy cluster is given by
\begin{equation}
\label{dl}
 D_{L}(z)=(1+z)c\int_0^z\frac{dz'}{H(z')},
 \end{equation}
 where $c$ is the speed of light and
  \begin{equation}
	\label{hz}
  H(z)=H_0 \sqrt{\Omega_M(1+z)^3+(1-\Omega_M)}.
  \end{equation}
Here, $\Omega_M = 1-\Omega_\Lambda$, where $\Omega_\Lambda$ is the cosmological constant density parameter and $H_0$ is the Hubble constant. Besides  priors quoted previously, we also use $H_0=67.27 \pm 0.66$ km s$^{-1}$ Mpc$^{-1}$.  As used in previous works that explored the depletion factor \cite{bat,pla}, we consider the following function for $\gamma(z)$ to explore a possible redshift evolution: $\gamma(z)=\gamma_0(1+\gamma_1z)$. 

Our results are in Fig.(2). In fig.(2a) we plot the 1$\sigma$, 2$\sigma$ and 3$\sigma$ c.l. regions for $\gamma_0$ and $\gamma_1$. We obtain at 1$\sigma$: $\gamma_0=0.86 \pm 0.04$ and $\gamma_1=-0.04 \pm 0.12$. These results are in full agreement with those from SNe Ia data. In Fig.(2b) we plot the evolution of the function $\gamma(z)$ and its 1$\sigma$ interval (dashed line).  As one may see, these results also support no redshift evolution for the depletion factor in galaxy clusters. The constraints are tighter at low redshift than those from the GP regression, which reflects the adopted parameterization.}

\begin{table}
\caption{Constraints on the gas depletion factor obtained from GP reconstruction method at different redshifts using the full sample of the Union2.1 compilation and 31 binned data from JLA compilation. We also write the results from the analysis by using luminosity distances  obtained from the flat $\Lambda$CDM model constrained by the current cosmic microwave background radiation observations. For this case, we consider a simple function for $\gamma(z)$, $\gamma(z) = \gamma_0(1+\gamma_1 z)$. The error bars correspond to 68.3\% C.L.}
\label{tables1}
\par
\begin{center}
\begin{tabular}{|c|c|c|c|c|c|}
\hline\hline  \quad Sample \quad & $z=0.0$  \quad & $z=0.5$  \quad & $z=1.0$ \quad 
\\ \hline\hline 
\hline 
{Union2.1} &  \quad $0.85 \pm 0.09$ \quad &  \quad $0.85 \pm 0.09$ \quad &  \quad $0.81 \pm 0.10$ \quad \\
\hline
\hline
JLA &  \quad $0.81 \pm 0.08$ \quad &  \quad $0.83 \pm 0.08$ \quad &  \quad $0.83 \pm 0.08$ \quad \\
\hline
\hline
$\Lambda$CDM Planck&  \quad $0.85 \pm 0.06$ \quad &  \quad $0.83 \pm 0.10$ \quad &  \quad $0.80 \pm 0.15$ \quad \\
\hline \hline
\end{tabular}
\end{center}
\end{table}

\section{Discussion}

{  How can we interpret our results and what are their usefulness? First of all, let us remind ourselves of the standard assumptions regarding cosmological analyses using the gas mass fraction measurements. First, a prior for $\gamma_0$ is adopted based on simulations within a given very specific cosmological model. Second, as there is no physically motivated model for a redshift evolution of $\gamma$, a simple linear parameterization is adopted. Our method can remove these limitations since i) we use data directly, no need to adopt a specific cosmological model, no use of simulations and ii) the lack of a physically motivated function for $\gamma(z)$ is dealt by considering a Gaussian Process reconstruction, which does not assume a given parameterization thereby letting the data decide. The small dependence on a cosmological model from SNe Ia data is not an issue for our method. SNe Ia can in principle be calibrated with $H(z)$ from cosmic chronometers \cite{javier} or baryon acoustic oscillation \cite{jahar}.}

Our results point that the approximations done so far are reasonable, since our constraints cover essentially the range provided by simulations. Thus, our method validates the previous results from the literature, since there is no indication of a discrepancy of our constraints for $\gamma(z)$ and the values for the simulations. In addition,  our constraints can be used as a prior for tests of alternative cosmological models using the gas mass fraction. This is really an advantage given that simulations so far are available only for a specific $\Lambda$CDM framework ($\Omega_M=0.24$ and $\Omega_{\Lambda}=0.76$ and $H_0=72$ km/s/Mpc) and are very costly to obtain. By inspecting Eq. \ref{gamma}, we can see there is a clear degeneracy between $\gamma$ and $\Omega_M$, so for models where dark matter and baryons evolve in the same fashion with redshift our results should be scaled by a constant factor only. When that is not the case, as for example for a model where dark energy interacts with dark matter, the interpretation becomes trickier since one cannot disentangle dark matter evolution from $\gamma$ evolution.

Finally, we detected that our constraints are currently limited by systematic errors as the mass calibration of clusters or the power-law index of the density profiles. Therefore, greater samples will be useful only if efforts to mitigate those effects are undertaken.

\section{Conclusions}

Galaxy clusters are the largest gravitationally collapsed objects in the universe, which makes them especially interesting for cosmology. In particular, measurements of the gas mass fraction in galaxy clusters have been used as an independent cosmological probe, posing increasingly tighter constraints on the main cosmological parameters and on gravity theories \cite{allen4}. A crucial assumption in this kind of analysis is the constancy of $f_{gas}$ with redshift, which is measured by the X-ray emitting gas depletion factor $\gamma$, i.e., the ratio by which the hot gas fraction in galaxy clusters is depleted with respect to the universal mean.

In this work, differently from previous studies, in which a possible evolution of this  quantity has been tested through  hydrodynamical simulations based on a specific cosmology, we  have used only cosmological observations, i.e., 40 measurements of the gas mass fraction and 580 distance measurements of SNe Ia, along with the validity of the CDDR, to reconstruct the evolution of $\gamma$ up to $z = 1$. This reconstruction was performed by using Gaussian Processes. We have also performed an analysis by using the 31 binned distance moduli of SNe Ia from the JLA compilation. As shown in Table I, the intervals of values for the gas depletion factor obtained in our analysis as well as the evidence of no evolution with redshift not only strongly support the results from cosmological hydrodynamical simulations \cite{pla,bat} but also corroborate the arguments behind the analyses using the gas mass fraction as a cosmological probe. {  More to the point, the results were completely consistent with those assuming a flat $\Lambda$CDM model for the luminosity distances constrained by CMB observations. Our results not only validate the standard approach but also provide a prior on $\gamma(z)$ for cosmological analyses using different cosmological models.} 


Finally, it is worth mentioning that when larger $f_{gas}$ samples (mainly at high redshifts) with smaller statistical and systematic uncertainties become available, more robust analyses of the type proposed here will either corroborate or even contradict the results of the hydrodynamical simulations. {  More specifically, our results have shown that progress towards this direction depends strongly on the mass calibration and gas profile measurements, which are the current limiting factor of our method.} However, with those errors under control, different results from our method and  the hydrodynamical simulations may indicate the presence of some  unknown mechanism in the intra-cluster medium not yet modeled in the simulations.

\vspace{1cm}

\section{Acknowledgments}

RFLH acknowledges financial support from CNPq/Brazil (No. 303734/2014-0). JGS is supported by CAPES/Brazil and FAPERJ (Rio de Janeiro State Research  Foundation). VCB is supported by S\~ao Paulo Research Foundation (FAPESP)/CAPES agreement under grant number 2014/21098-1 and FAPESP under grant 2016/17271-5. JSA is supported by CNPq/Brazil and FAPERJ. F.A-S. acknowledges support from {\it Chandra} grant GO3-14131X.

\label{lastpage}
\end{document}